\documentclass[a4paper,12pt]{article} 

\usepackage{cite}
\usepackage[nohead,margin=1in]{geometry}
\usepackage{color}
\usepackage{graphicx}
\usepackage{amsmath,amssymb}
\usepackage[colorlinks=true,bookmarks=false,citecolor=blue,urlcolor=blue]{hyperref}

\usepackage{mathptmx,courier,textcomp}
\usepackage[scaled=.92]{helvet}

\newcommand\MEETtitle[1]{{ \noindent \Large \bf \begin{center} #1 \end{center}\rm } \vskip.1in \rm\normalsize }

\newcommand\MEETauthor[1]{\hskip2.5pc \parbox{.8\textwidth}{ \noindent%
		\normalsize \bf \begin{center} #1 \end{center}\rm } \vskip-1pc }

\let\title\MEETtitle
\let\author\MEETauthor

\let\address\MEETaddress

\begin{document}

\title{Ultra-high Sensitive Surface Plasmon Resonance Based Sensor with Dual Resonance}

\author{Indrajeet Kumar$^1$, Ranjeet Dwivedi$^2$, Saurabh Mani Tripathi$^{1,2}$}

\address{$^1$Centre for Lasers and Photonics, Indian Institute of Technology Kanpur, Kanpur 208016, India\\
$^2$Department of Physics, Indian Institute of Technology Kanpur, Kanpur 208016, India}

\vskip2pc
\noindent
\textbf{Abstract:} We show that for the optimized angle of incidence, the SPR based optical sensors exhibit dual resonance in the near-infrared region around which the sensor becomes exceptionally high sensitive. Both the resonances show opposite spectral shift with an ambient refractive index, increasing the differential shift by many folds. The physical reason behind the dual resonance and opposite spectral shift are also explained using the modal analysis and the phase-matching condition. The presence of dual resonance is highly susceptible to the angle of incidence, facilitates the sensors to work in a wide range from gaseous specimen to biological ones with extremely high sensitivity of 460 $\mu m/RIU$ and 290 $\mu m/RIU$, respectively. The considered sensor has broader prospects for the detection of bio-chemicals and gases without changing its geometrical parameters.

\section{Introduction}

Surface Plasmon resonance (SPR) is one of the most promising optical techniques for sensing applications due to their high sensitivity, label-free detection as well as versatility in various fields such as physical \cite{Wang, Taylor}, chemical, \cite{Homola} and bio-molecules \cite{Homola, Jatschka, Tomyshev}. In this technique, the electron density oscillations at the metal-dielectric interface are excited by the \textit{\textit{p}}-polarized light at a specific angle/wavelength. The matching of the momentum of incident light to that of those oscillations leads to the resonance \cite{KC}. The electromagnetic field near the metal-dielectric interface is strongly localized and therefore, any changes in the refractive index of the surrounding medium alters the resonance condition making SPR highly sensitive \cite{Yu}. In order to achieve the SPR, various configurations have been widely reported such as prism based Kretschmann configuration \cite{KC}, metallic gratings \cite{Roh}, directional couplers based on dielectric waveguide to metal strip coupling, metal coated waveguides and fibers. SPR sensors based on Krestman configuration have been widely used for sensing of liquids, gases, bio-chemicals etc. The refractive index (RI) sensitivity of such sensor in the visible region is $\sim$ 5 $\mu m/RIU$. In order to enhance the sensitivity several attempts were made by researchers such as the integration of two-dimensional materials like graphene, tungsten disulfide ($WS_2$), molybdenum selenium ($MoSe_2$) with metallic thin films \cite{Luo, Wang1, Liu, Patil, Xu}. Luo \textit{et al.} investigated the atomically thin layer of $MoSe_2$ on gold thin film and reported 36.3 $\%$ increased sensitivity compared to standard gold film SPR \cite{Luo}. Wang \textit{et al.} reported the sensitivity enhancement by 26.6 $\%$ by modification of gold film with $WS_2$ nano-sheet overlayer \cite{Wang1}. Xu \textit{et al.} proposed a hybrid structure of graphene-aluminum-graphene which is capable of increasing the sensitivity by 3.4 times compared to aluminum based sensors \cite{Xu}.  Liu \textit{et al.} proposed silver-gold bimetallic film to enhance the sensitivity \cite{Liu} of a graphene–barium titanate-based SPR biosensor by 15 $\%$. Chabot et al. demonstrated sensitivity enhancement by 50 $\%$ using long-range surface Plasmon for toxicity monitoring with living cells \cite{Chabot}. Although these sensors give better sensing performance compared to single metal film, fabrication of such sensors requires a multi-step process. 

In this letter, we report that an ultra-high sensitivity can be realized by just optimizing the angle of incidence in Kretchmann configuration based RI sensor. At the optimized incident angle, the phase matching condition is satisfied for two different wavelengths in the NIR region, resulting in two highly sensitive resonance dips in the reflection spectrum. Furthermore, both the dips have opposite spectral shift with the ambient refractive index (ARI) change, enhancing the differential spectral shift and hence the RI sensitivity. Dual resonance with opposite spectral shift have been widely reported in modal interferometers and long period gratings in optical fibers and waveguides, which is due to the dispersion turning point (DTP) on either sides of which the group effective index difference of the participating modes has opposite sign \cite{SMT, GB, Li}. However, unlike DTP, in the considered structure the opposite spectral shift is due to the opposite dispersion behaviour of the phase matching condition at the two dips. 
Furthermore, the same sensing probe can be utilized to achieve ulta-high sensitivity for the detection of bio-chemicals as well as gases by just optimizing the angle of incidence. 

\section{Modelling}

\begin{figure}[b]
	\begin{center}
		\includegraphics [width=8 cm]{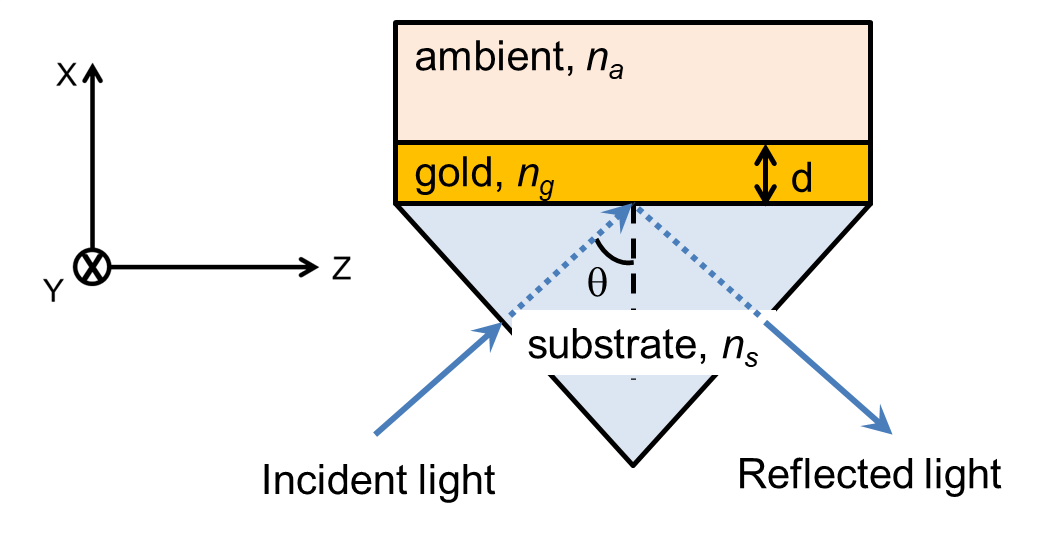}
	\end{center}
	\caption {Schematic diagram of the substrate/gold/ambient layered structure.}
	\label{schematic}
\end{figure}                 

For the study of SP behavior of the gold thin film, a layered structure is considered, which is shown schematically in Fig. \ref{schematic}. It consists of a thin gold film on a glass prism (substrate) and a sensing medium on its top. The refractive indices of the substrate, gold layer, and the analyte layer are denoted by $n_s$, $n_g$ and $n_a$, respectively. The thickness of the metal layer is denoted by $d$, which is taken as 45 nm. In the considered structure, the \textit{p}-polarized electromagnetic wave at an angle $\theta_i$ is launched from the substrate region, which excites the surface plasmon polariton (SPP) modes supported by the structure and when the phase matching condition is satisfied, i.e., the wave vector of the incident beam matches with the propagation constant of the modes, incident wave gets coupled to that SPP mode. The equation governing the phase matching condition is given by

\begin{equation}
n_sSin \theta_i = n_{eff}
\label{n_eff}
\end{equation}

where $n_{eff}$ is an effective refractive index of the SPP mode supported by the considered structure. Due to power coupling from incident wave to the SPP mode, a dip will appear in the reflection spectrum at the phase matching wavelength, for a fixed incident angle. With the change in ARI, the phase matching condition is satisfied at a different wavelength resulting in a shift in the reflection dip. The spectral shift in the reflection dip for a unit change in the ARI is termed as RI sensitivity. The reflection spectrum for the structure shown in Fig. \ref{schematic} can be obtained numerically using the following equation

\begin{equation}
R = \left|\frac{r_{sg}+r_{ga}exp(2ikd)}{1+r_{sg} r_{ga}exp(2ikd)} \right|^2
\label{reflectance}
\end{equation}

where, $r_{sg}$ and $r_{ga}$ are the reflection coefficient at the substrate-gold interface and gold-analyte interface, respectively, and \textit{k} is the propagation constant corresponds to a wavelength $\lambda$.

\section{Results and Discussion}
\begin{figure}[h!]
	\begin{center}
		\includegraphics [width=12 cm]{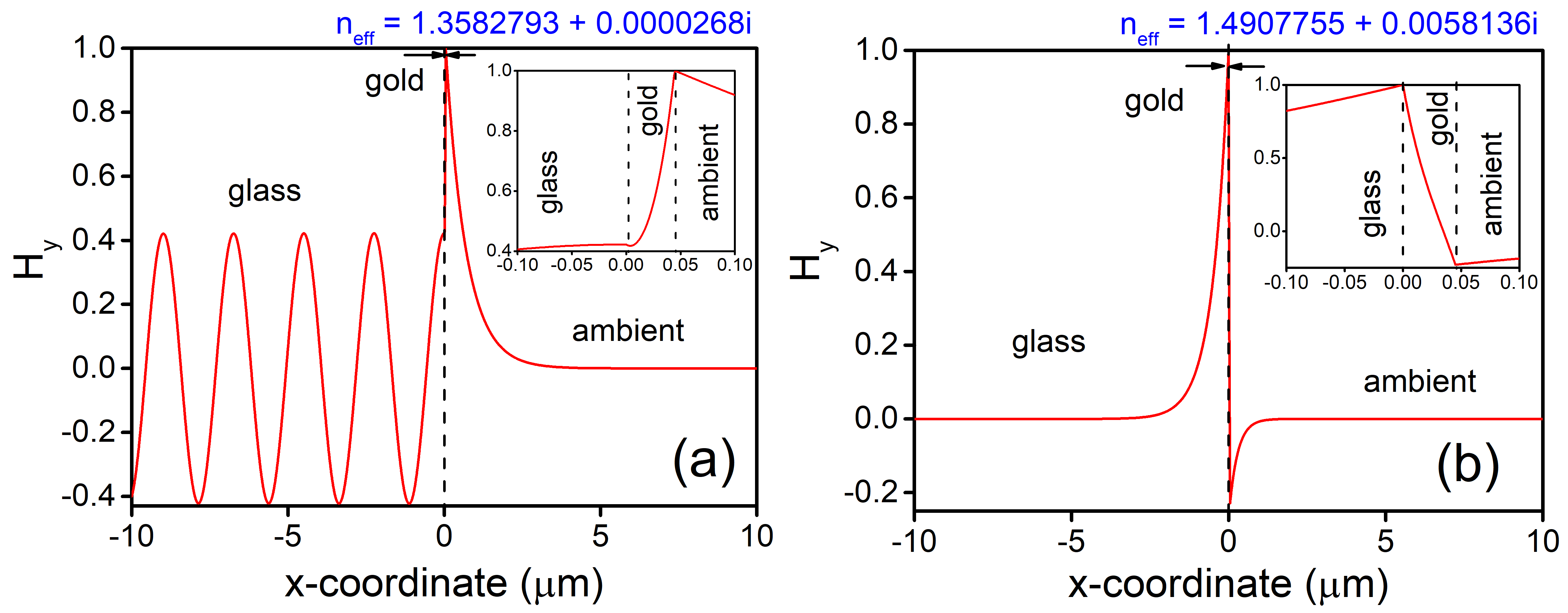}
	\end{center}
	\caption {Modal field distribution of magnetic field ($H_y$) for (a) leaky and (b) bound mode for wavelength 1.133 $\mu m$ and ambient refractive index of 1.330.}
	\label{1133nm_Hx}
\end{figure}

\begin{figure}
	\begin{center}
		\includegraphics [width=7 cm]{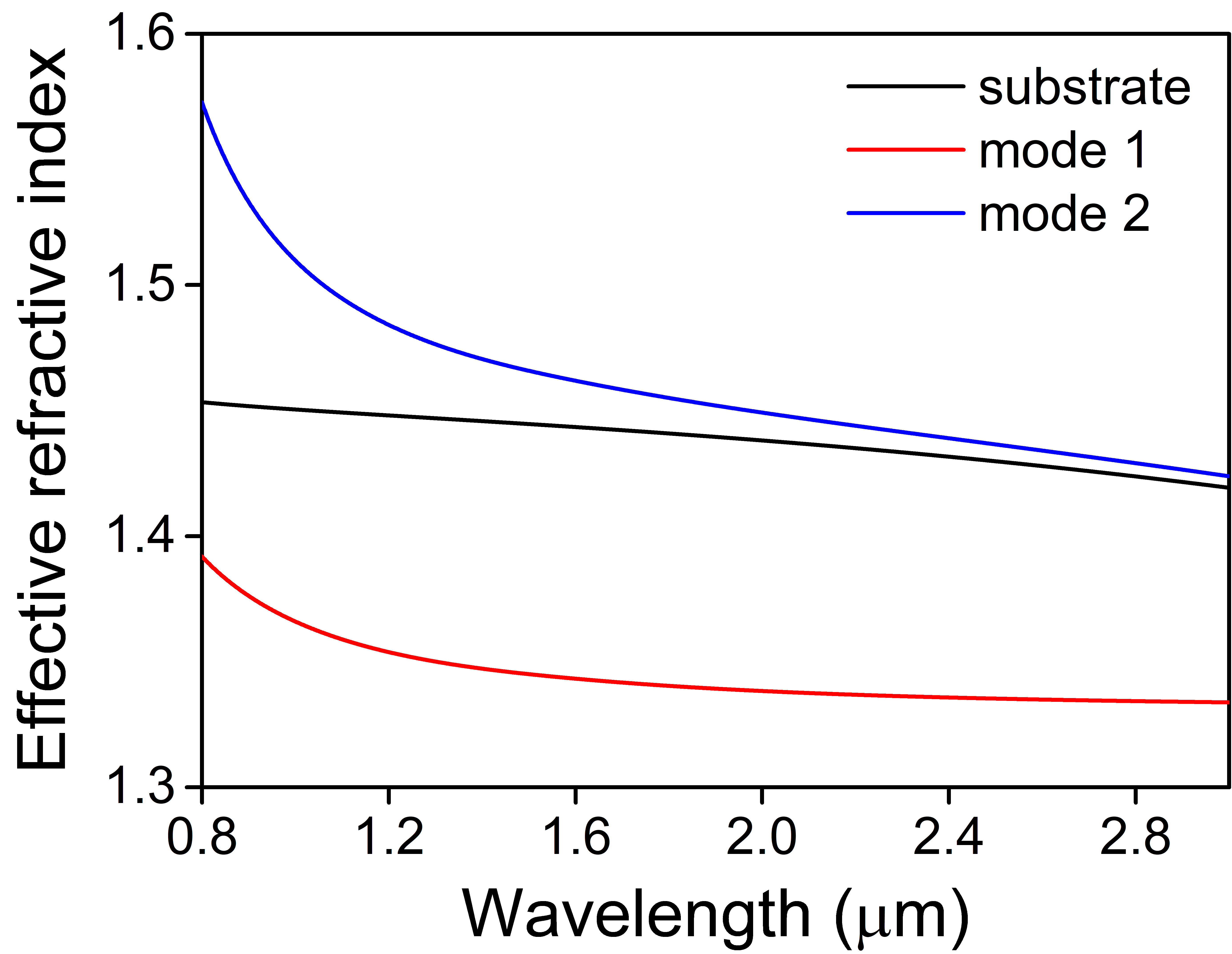}
	\end{center}
	\caption {Dispersion curve of the substrate refractive index and the modal effective refractive indices of the considered structure for an ambient refractive index of 1.330.}
	\label{modes}
\end{figure}

Since the phase matching condition and hence, the sensitivity depends on the interaction of the modal fields with ARI, we first analyze the modes supported by the structure shown in Fig. \ref{schematic}. The  magnetic field ($H_y$) distribution of the SPP modes supported by the structure at a wavelength of $\lambda$ = 1.133 $\mu m$ and $n_a$ = 1.330 are plotted in Fig. \ref{1133nm_Hx}. Here the modes are obtained by solving Maxwell’s equations using the method outlined in \cite{Ghatak2}. In our calculations, the wavelength dependence of the refractive index of fused silica is taken from the Sellmeier equation \cite{Ghatak}, and that of gold is taken form the Drude-Lorentz model \cite{Rakic}. The first mode (Fig. \ref{1133nm_Hx} (a)) having $(n_{eff})_R$ = 1.35827 is bounded in the ambient region and is oscillatory in the substrate whereas the second mode (Fig. \ref{1133nm_Hx} (b)) having $(n_{eff})_R$ = 1.49077 is bounded in both the substrate and ambient regions. Figure \ref{modes} shows the spectral variation of effective indices of both the SPP modes and the substrate RI. It may be noted that the $n_{eff}$ corresponding to the second mode (Fig. \ref{1133nm_Hx} (b)) is higher than the substrate RI, and hence the phase matching condition can't be satisfied for any angle of incidence. Furthermore, for a given wavelength, the phase matching condition can be satisfied for the first mode (Fig. \ref{1133nm_Hx} (a)) at an appropriate value of the angle of incidence.

\begin{figure}[h]
	\begin{center}
		\includegraphics [width=8 cm]{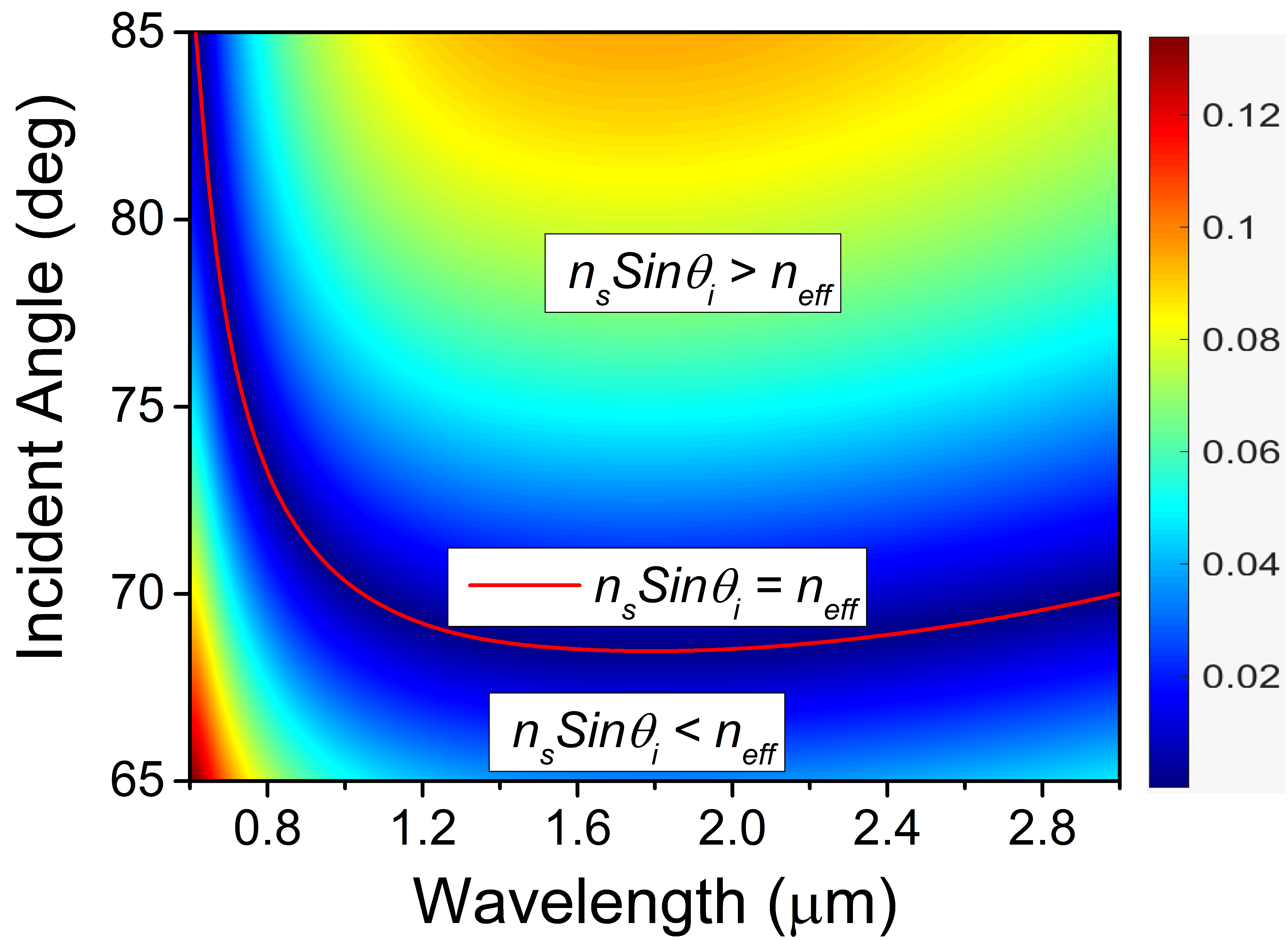}
	\end{center}
	\caption {Variation of the phase mismatch factor ($n_sSin \theta_i - n_{eff}$) with the incident angle and wavelength for the ambient refractive index of 1.330. Here solid red line corresponds to the phase matching points.}
	\label{angle_optimization}
\end{figure}

\begin{figure}
	\begin{center}
		\includegraphics [width=7 cm]{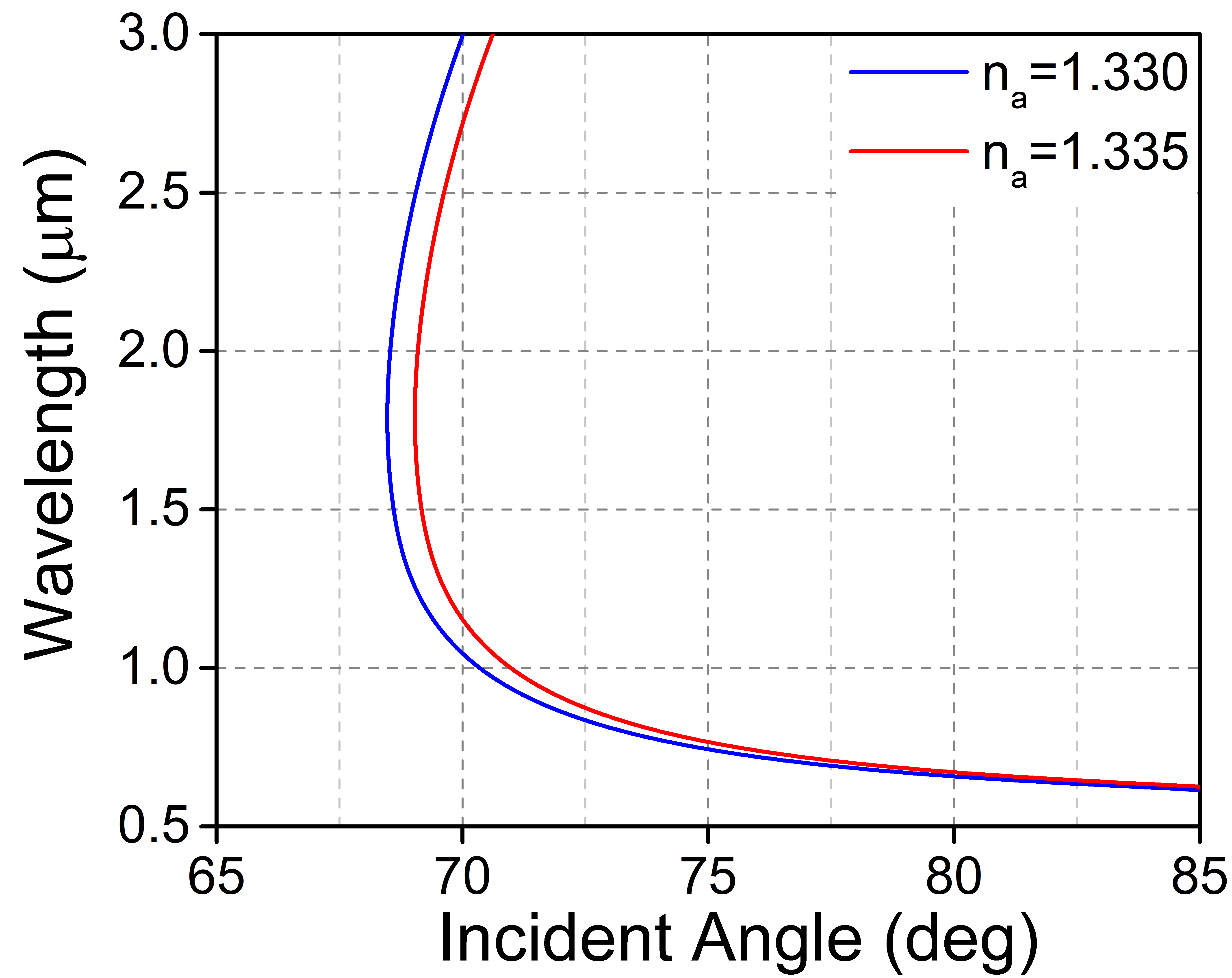}
	\end{center}
	\caption {Variation of the phase matching wavelength with the incident angle for two different ARI values 1.330 and 1.335.}
	\label{neff_2ARI}
\end{figure}

For a range of wavelengths, there is a minimum angle at which the momentum of the incident photon becomes equal to that of the layered structure and the resonance occurs, \textit{e.g.} for resonant wavelength to be in visible range the incident angle should be higher than 75$^\circ$, Fig \ref{angle_optimization}. This is a general case where only one resonant wavelength occurs. To study the dependence of the resonance behavior on the incident angle in detail, the phase matching condition is explored by varying the angle of incidence.
Figure \ref{angle_optimization} shows the variation of phase mismatch factor ($n_{eff} - n_sSin \theta_i$) with the incident wavelength and the angle of incidence. Here the incident angle is varied in the range 85-65$^\circ$ in an interval of 0.1$^\circ$ for the wavelength range 0.800-3.000 $\mu m$ and ARI of 1.330. At the incident angle of 85$^\circ$, single resonant position is observed at wavelength 0.630 $\mu m$ and shifts towards the longer wavelength with a decrease in the incident angle. It also can be observed that at a specific point, the Eq. \ref{n_eff} is satisfied at two wavelengths. For example, at an incidence angle of 69.5$^\circ$ the phase matching wavelengths corresponds to 1.133 and 2.755 $\mu m$. These two wavelengths shift opposite with the further decrease in the incident angle and converge to a single one at a certain angle. The variation in phase matching wavelength with ambient is also studied and plotted in Fig. \ref{neff_2ARI} for two ARI; 1.330 and 1.335. It may be noted that in the NIR region, for fixed angle, the shift in phase matching wavelength with ARI change is much higher than that in the visible region and hence very high sensitivity is expected in this region.

\begin{figure}
	\begin{center}
		\includegraphics [width=8 cm]{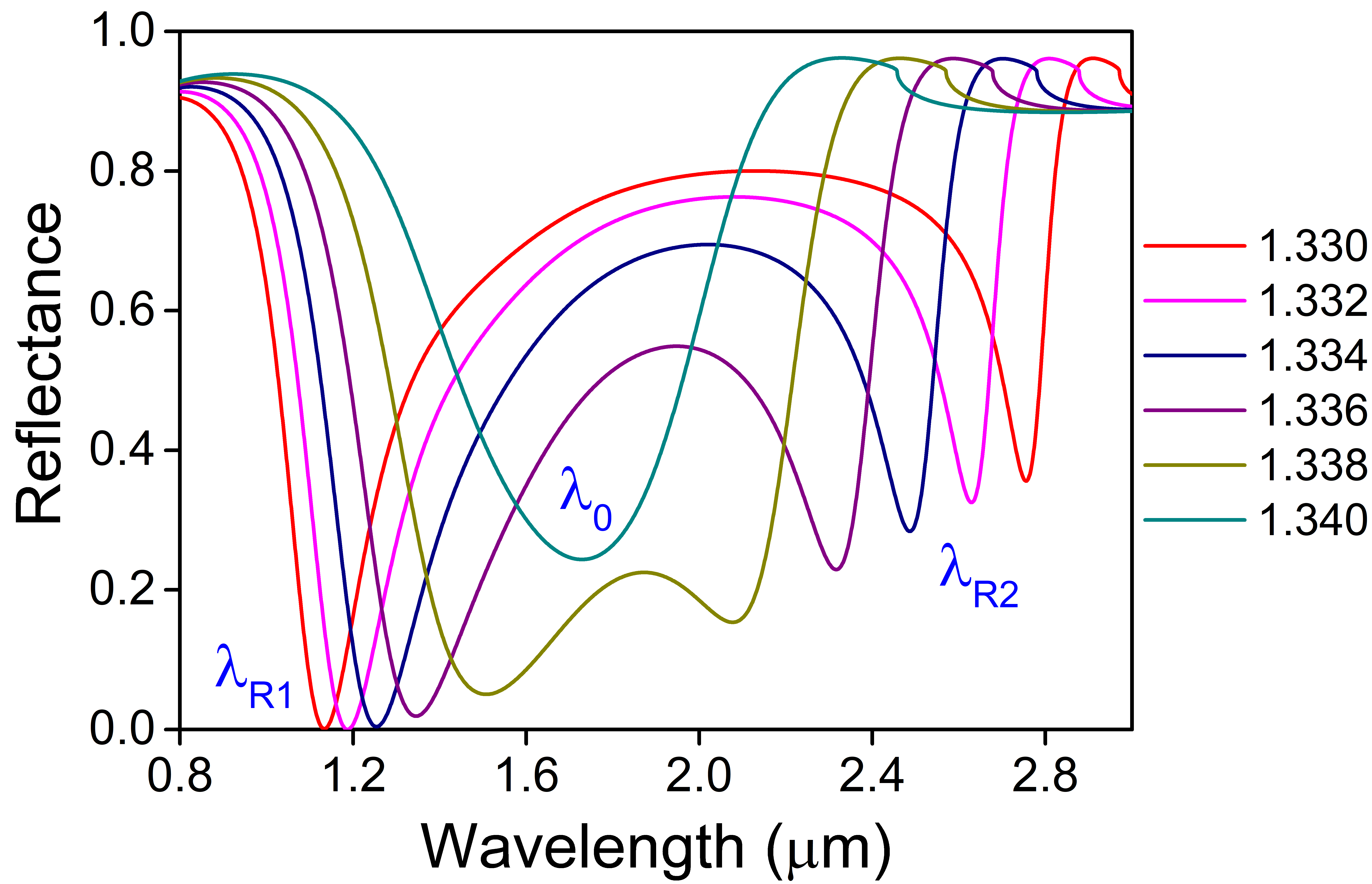}
	\end{center}
	\caption {Reflection spectra of considered structure at the incident angle of 69.5$^\circ$ for the ambient refractive index range 1.330-1.340.}
	\label{reflection}
\end{figure}

In order to estimate the sensitivity of the considered sensor structure, we have shown the reflection spectrum (calculated using Eq. \ref{reflectance}) in Fig. \ref{reflection} for different ARI values. Here the incidence angle is selected as 69.5$^\circ$ such that the phase matching is satisfied for two different wavelengths near the turning point simultaneously, in the ARI range 1.330-1.340. The presence of dual resonance can be seen clearly from the reflection spectrum, appearing as two spectral dips having opposite shift, similar to those reported in the case of long-period fiber grating \cite{SMT}.

\begin{figure}
	\begin{center}
		\includegraphics [width=6 cm]{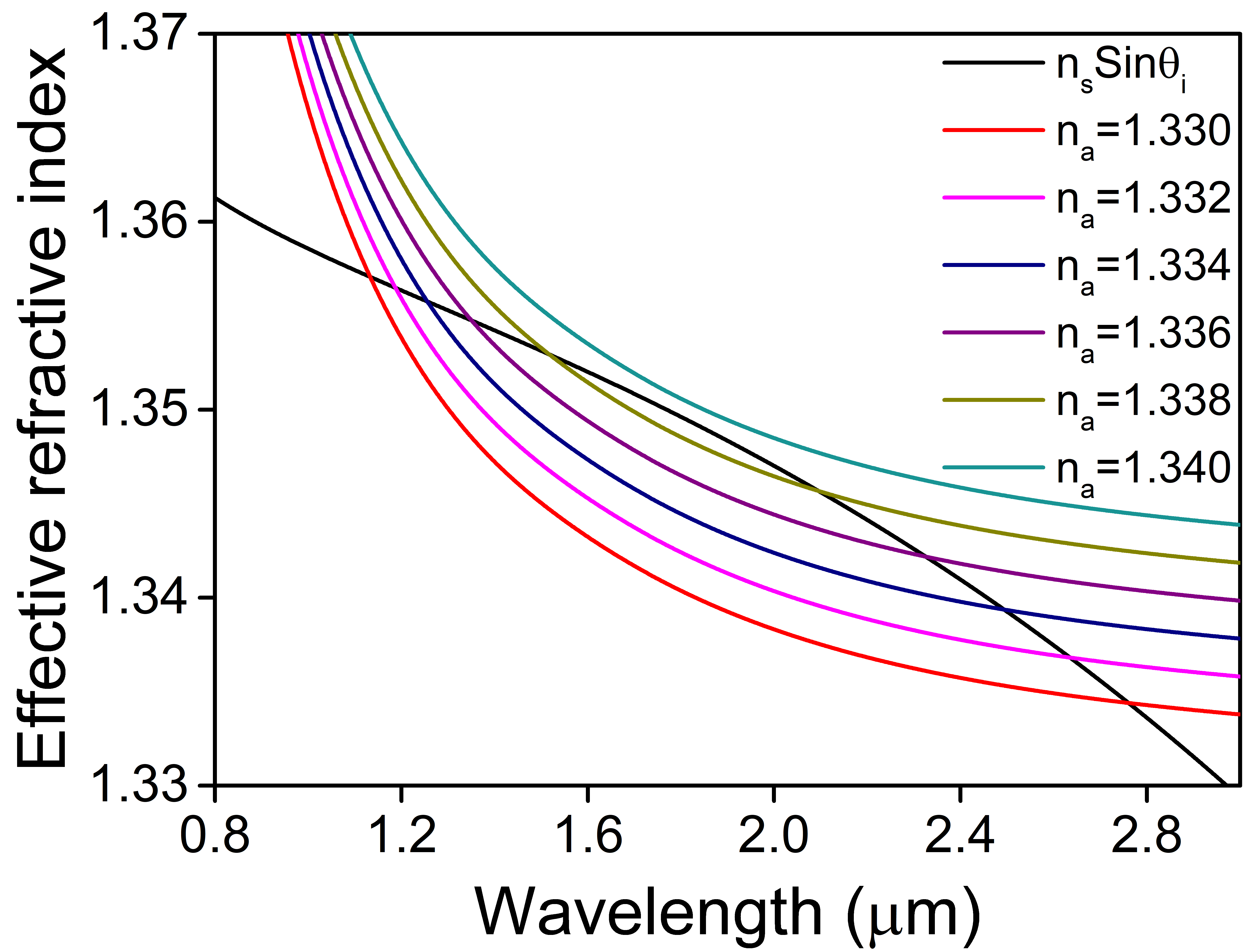}
	\end{center}
	\caption {Phase matching curve for ambient refractive index range 1.330-1.340 with incident angle of 69.5$^\circ$.}
	\label{neff}
\end{figure}

For convenience, the resonant position in lower wavelength region is denoted as $\lambda_{R1}$ and that in higher wavelength region is $\lambda_{R2}$. For $n_a$ = 1.330, the resonance appears at 1.133 and 2.755 $\mu m$, and with an increase in the refractive index of the surrounding, it is observed that $\lambda_{R1}$ is red-shifted whereas $\lambda_{R2}$ is blue-shifted. These two resonance wavelengths converse to a single one ($\lambda_0$) at 1.729 $\mu m$ for the refractive index of 1.340, which corresponds to the \textit{Turning Point}. From Fig. \ref{reflection}, it is also observed that for a change in ARI, the spectral shift in resonant position near the \textit{Turning Point} is maximum. The opposite shift in resonance dips is due to the opposite shift in the phase matching wavelengths, which can be observed clearly in Fig. \ref{neff}. Here the intersection points correspond to the phase matching / resonance positions. The opposite shift in resonant wavelengths provides the increased differential spectral shift as compared to the single resonant dip, leading to enhanced sensitivity which is given by 

\begin{equation}
S = \left| \dfrac{\Delta (\lambda_{R1}-\lambda_{R2})}{\Delta n_{a}}\right|  
\label{1}
\end{equation}

\begin{figure}[h]
	\begin{center}
		\includegraphics [width=7 cm]{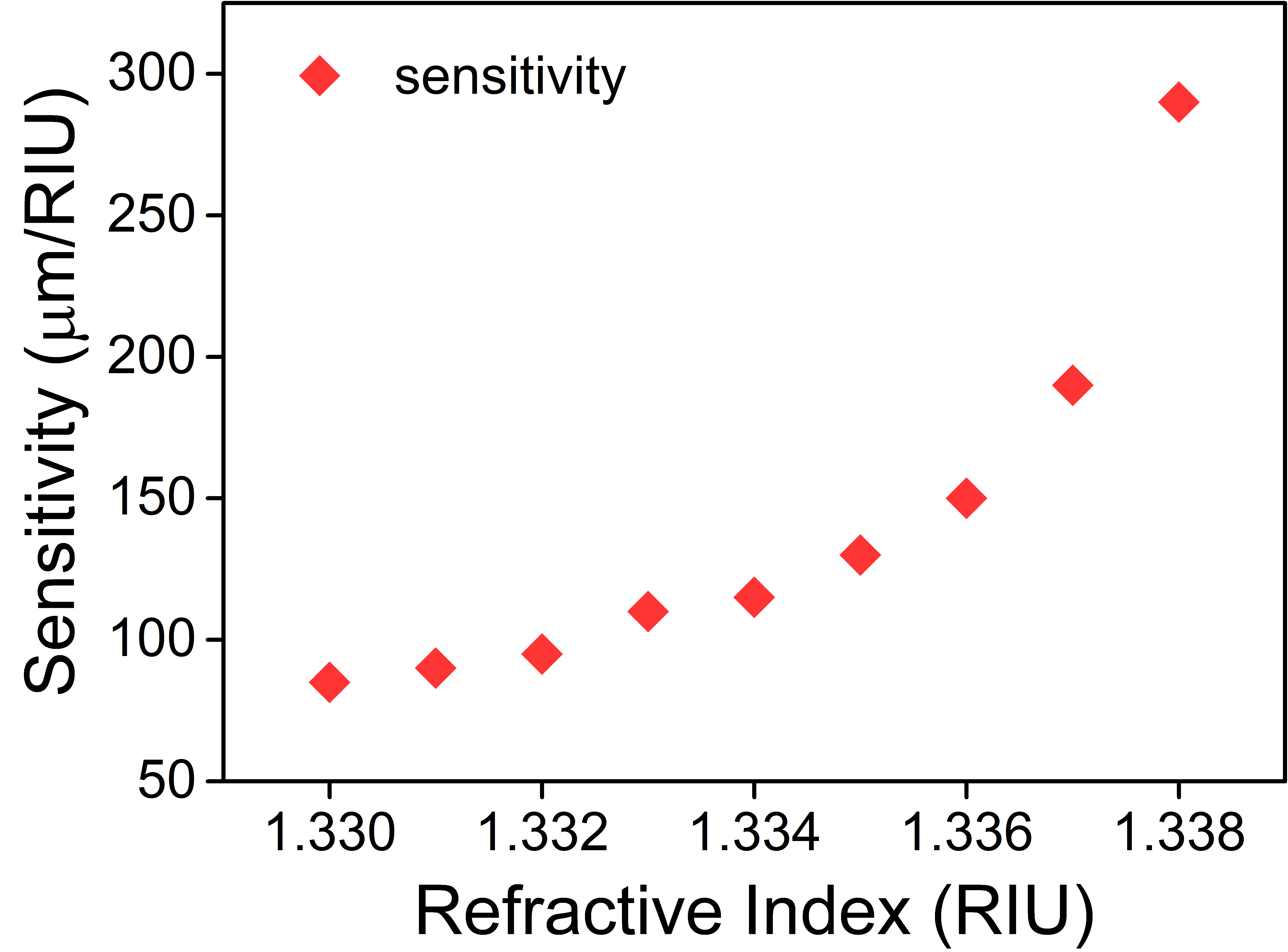}
	\end{center}
	\caption {Variation of the sensitivity for ambient refractive index $n_a \pm 0.0001$ and incident angle of 69.5 $^\circ$.}
	\label{sensitivity}
\end{figure}

\begin{figure}[h]
	\begin{center}
		\includegraphics [width=7 cm]{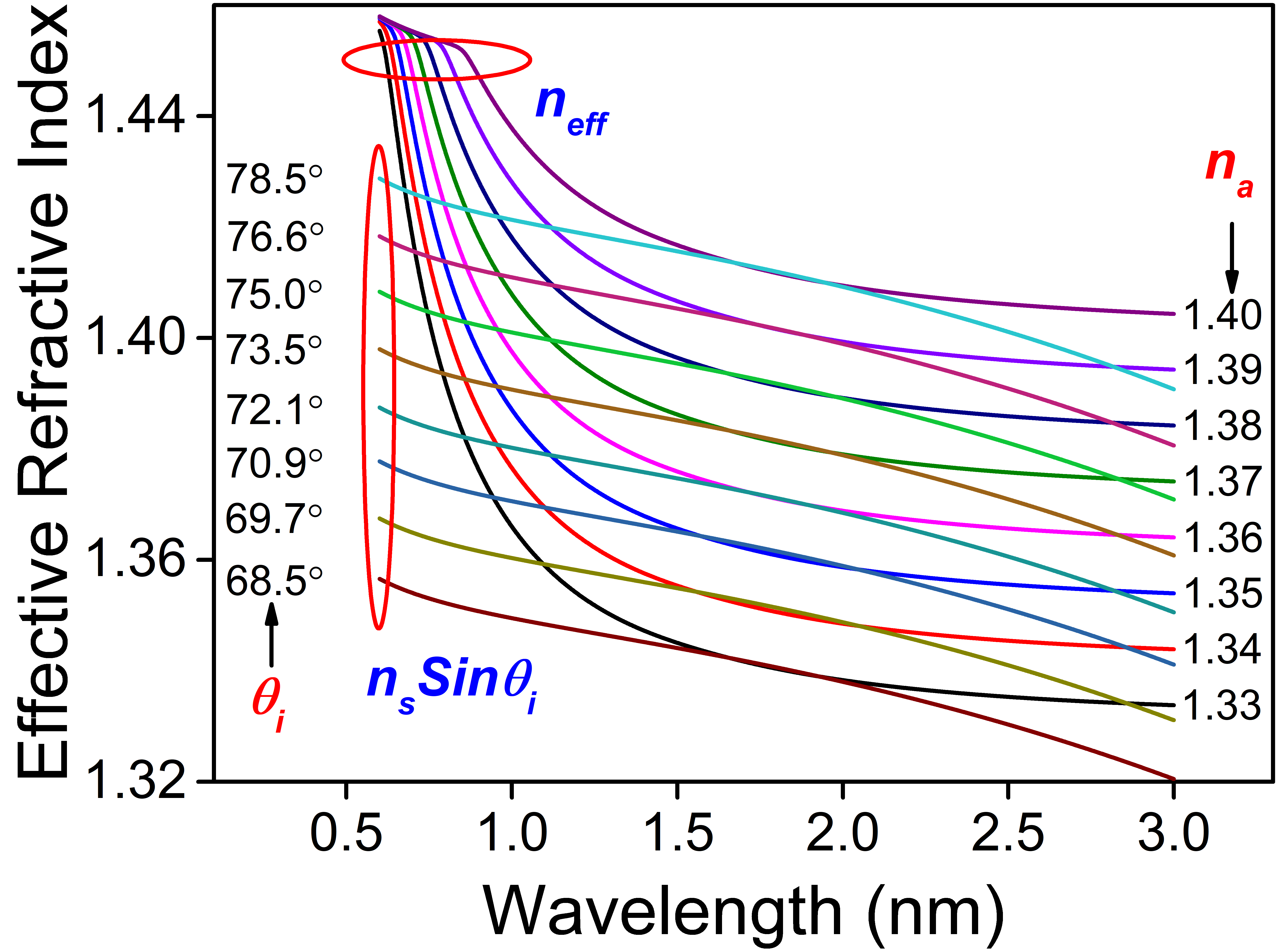}
	\end{center}
	\caption {Phase matching curve for ambient refractive index,$n_a$, ranging 1.330-1.400 and  incident photons at various angles. Here the angle is optimized to get the optimum value of sensitivity for each $n_a$, in the dual resonance.}
	\label{DR}
\end{figure}

The sensitivity calculated at the incident angle of 69.5 $^\circ$  for an $n_a$ varying from 1.330 to 1.338 is plotted in Fig. \ref{sensitivity}, considering ARI change $\pm$ 0.0001 at each $n_a$. The maximum sensitivity, 290 $\mu m/RIU$, is obtained around ARI 1.3380, for which the resonance dip is closest to the \textit{Turning point}. The sensitivity is observed to be decreased for the resonance dip appearing away from the \textit{Turning Point} and obtained to be 85 $\mu m/RIU$ for ARI $1.330$.
However, similar sensitivity can be obtained for all values of $n_a$ if we get \textit{Turning point} corresponding to that $n_a$. The tuning of the incident angle solves this issue. Figure \ref{DR} shows the phase matching curve for ARI range 1.330-1.400. It is observed that there is always an incident angle for which every refractive index has a \textit{Turning point}. An expression is derived using polynomial fit in order to obtain an incident angle corresponds to the \textit{Turning point} for $n_a$ in range 1.33-1.40 and given by
\begin{equation}
\theta_T = An_a^3 + Bn_a^2 + Cn_a + D
\label{angle_turning_bio}
\end{equation}
where, $\theta_T$ is the angle of incidence corresponding to the \textit{Turning Point} for $n_a$ in the range 1.330-1.400, $ A = 5303.03 $, $ B = -21114.72 $, $ C = 28137.37 $, and $ D = -12480.40 $.

\begin{figure}
	\begin{center}
		\includegraphics [width=8 cm]{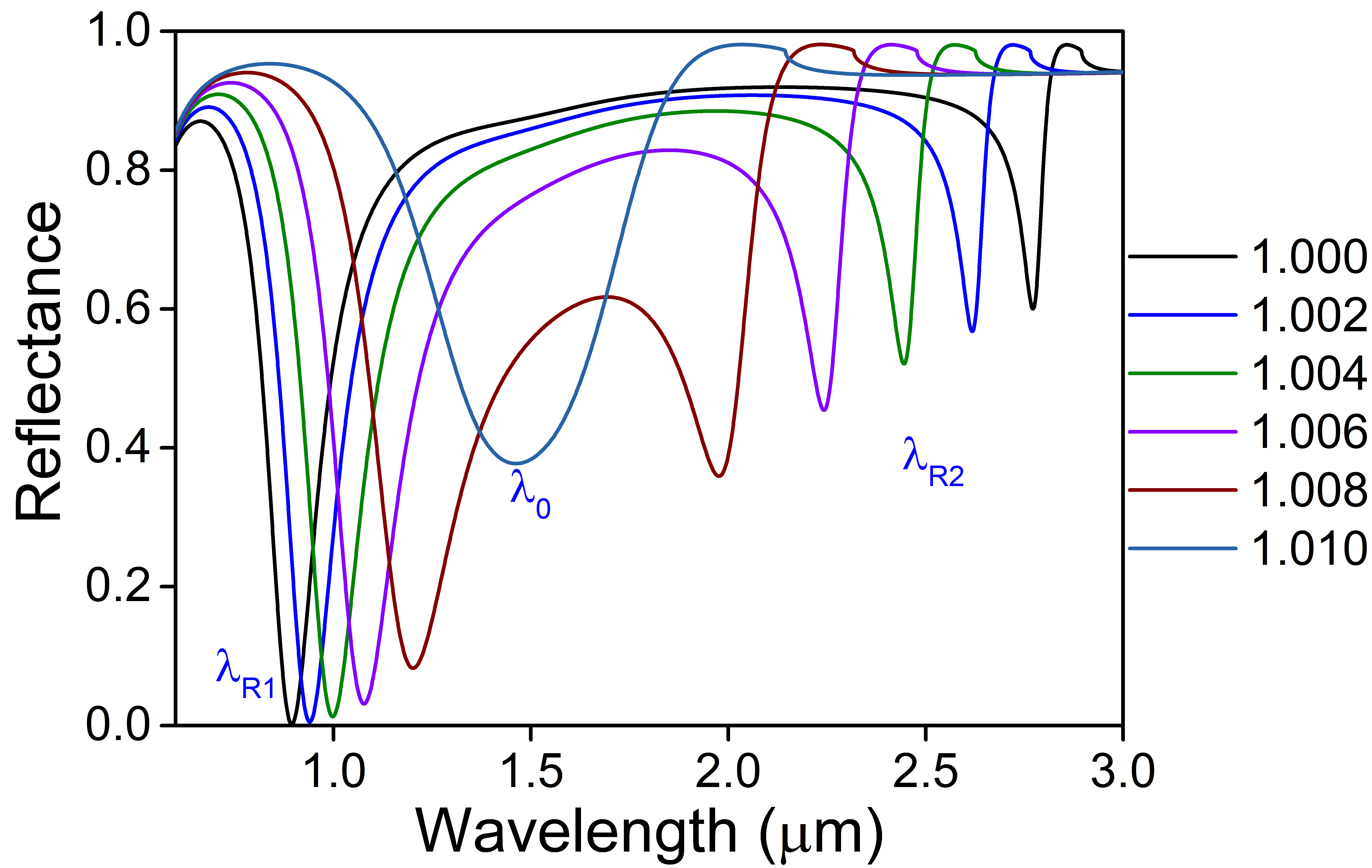}
	\end{center}
	\caption {Reflection spectra at the optimized incident angle of  44.7 $^\circ$ with ambient refractive index varying from 1.000 to 1.010.}
	\label{ref_gas}
\end{figure}

The idea of obtaining the \textit{Turning Point} for a wide range of refractive index in biological applications (1.330-1.400) is extended to gaseous RI range also. It is observed that the considered sensor with the same geometrical parameters possesses \textit{Turning Point} in a gaseous refractive index regime also but at a much lower incident angle. For the RI range 1.001-1009, the incident angle is optimized to be 44.7$^\circ$, which has all the features as discussed above for the RI range in biological applications. The calculated reflectivity at the incident angle of 44.7 $^\circ$ is plotted in Fig. \ref{ref_gas}. The sensitivity is obtained to be higher compared to that in biological RI range, which is found to be varying from 95 $\mu m/RIU$ and 460 $\mu m/RIU$ for ARI $1.000 \pm 0.0001$ and $1.009 \pm 0.0001$, respectively. In order to obtain the maximum sensitivity for the whole range of refractive index in the gaseous regime, the incident angle should be closest to the \textit{Tuning Point}, for which an expression is derived and given by
\begin{equation}
\phi_T = Pn_a + Q
\label{angle_turning_gas}
\end{equation}
where, $\phi_T$ is the angle of incidence corresponding to \textit{Turning Point} for $n_a$ in the range 1.100-1.200, $ P = 63.5519 $, and $ Q = -19.6419 $.

For some bio-applications, measurand molecules bind to the surface to the extent of few nanometers, which changes the refractive index at the surface, causing the shift in resonance dip. This shift facilitates the sensing of the measurand specimen. For the calculation of surface sensitivity, considering a protein layer having a refractive index of 1.45, the spectral shift is found to be 38 nm to 56 nm for the binding thickness of 1 nm to 5 nm, respectively. 

\section{Conclusion}
In conclusion, we have shown the existence of \textit{Turning point} in the phase matching condition for surface plasmon in gold thin film by optimizing the incident angle. Around the \textit{Turning point}, the phase matching condition is satisfied for two different wavelengths in the NIR region resulting in dual resonance dip in the reflection spectrum. Both the dips have opposite spectral shift with the ARI change, thereby increasing the differential shift and hence the sensitivity. The sensitivity is maximum closest to the \textit{Turning point}, the position of which can be tailored for different RI range. For the optimized parameters, the maximum sensitivity of the sensor is calculated to be 290 $\mu m/RIU$ for ARI range 1.330-1.338 and 460 $\mu m/RIU$ for ARI range 1.000-1.010. The considered sensor is easy to fabricate and versatile for measuring the sensitivity of biological as well as gaseous specimen with ultra-high sensitivity only by changing the angle of incidence. The surface sensitivity is also found to be very high 38-56 nm/nm for the binding thickness 1-5 nm.

\section{Acknowledgement}
The work was financially supported by the Science and Engineering Research Board, Government of India through projects PDF/2017/002679 and EMR/2016/007936.

\end{document}